# VARIABILITY OF CRITICAL ROTATIONAL SPEEDS OF GEARBOX INDUCED BY MISALIGNEMENT AND MANUFACTURING ERRORS

∗ Nicolas DRIOT

Emmanuel RIGAUD
Joël PERRET-LIAUDET

*Laboratoire de Tribologie et Dynamique des Systèmes*
*Ecole Centrale de Lyon*
*36 avenue Guy de Collongue B.P. 163*
*69131 Ecully cedex, FRANCE*
E-mail : nicolas.driot@ec-lyon.fr
E-mail : emmanuel.rigaud@ec-lyon.fr
E-mail : joel.perret-liaudet@ec-lyon.fr

**Keywords** : Gear, Tolerances, Statistical Moments, Critical Speeds

## ABSTRACT

Noise measurement on a population of gearbox manufactured in large number often reveals high variability due to tolerances on each gear design parameter (manufacturing errors). Gearbox noise results mainly from housing vibration induced by the gear meshing process. High acoustic levels correspond to excitation in a resonant manner of some critical modes. All gearbox modes and especially these critical modes depend on tolerances related to gear design parameters through the meshing stiffness.

Tolerances are considered on shaft misalignment errors, teeth profile errors and teeth longitudinal errors. We introduce these tolerances as geometric random gaussian parameters. The critical modes are extracted using an efficient procedure. The retained "statistical" method to treat tolerances is the Taguchi's method. We illustrate this methodology considering two different gearbox dynamic models fitted out with a single spur gear. Obtained dispersion results are validated by comparison with Monte-Carlo simulations.

## INTRODUCTION

The vibratory and acoustical behaviour of gearboxes results from numerous sources. Among these, it is generally admitted that the main source is the static transmission error under load (STE) [1]. STE is mainly governed by periodic components at the meshing frequency due to (1) elastic deflections of gear teeth under load (periodic meshing stiffness) and (2) teeth geometry modifications, manufacturing errors and shaft misalignments. Under operating conditions, STE generates dynamic mesh forces leading to dynamic forces and moments transmitted through bearings, housing vibration and noise.

Further, critical rotation speeds associated with high dynamic mesh forces and high noise levels, correspond to the excitation of some critical modes having a high potential energy stored by the meshing stiffness [2]. These critical speeds are mainly controlled by the time-average meshing stiffness.

At last, considering gearbox manufactured in large number, we observe dispersion of critical speeds and excitation levels mainly due to variability of STE and meshing stiffness. Sources of dispersion result mainly from geometry faults authorised by designers who introduce necessary tolerances. We model these uncertain parameters (geometry faults) by random parameters.

In this context, the aim of this paper is to deal with some results about variabilities of peak to peak STE, time-average meshing stiffness and critical speeds. Two different dynamic gearboxes are investigated to show efficiency of used process. Statistics are obtained from Taguchi's method and are compared to Monte-Carlo simulations. Sources of variability are teeth profile manufacturing errors, teeth longitudinal manufacturing errors and shaft misalignments.

## BRIEF DESCRIPTION OF STATISTICAL METHODS

### Monte-Carlo method

Well-known Monte-Carlo simulations are generally used to obtain reference predictions in order to test other methods. Requirement of a large number of samples increases strongly the computation time. In this study, we used 30000 samples. Further, the accuracy is also determined by the selected random generator. In order to generate Gaussian variables, we have used a Box and Muller algorithm.

### Taguchi's method

Taguchi's method allows to estimate in a very simple way the statistical moments of a function of multiple random variables whose probability densities are known [3]. Statistical moments are estimated from numerical integration of Gauss quadrature type. Then, the response function is calculated for a relatively short number of samples, judiciously chosen. For each uncertain variable, a number of samples up to three is necessary to take into account the eventual non-linear behaviour of the response function. Precision increases rapidly with this number of samples. In this study, we used 81 samples to treat four random parameters. The principal advantages of this method are the ease of its numerical implementation and its short computing time. Unfortunately, Taguchi's method can't provide response function probability density function (PDF).

## FIRST GEARBOX DYNAMIC MODEL

In this section, we defined all the steps required for the computation of critical speeds variability (STE statistics, meshing stiffness statistics, critical modes variability) and we present results associated to a simple gearbox dynamic model.

### Description of gear teeth geometry

A single-stage involute spur gear has been investigated (Table 1). Profile and longitudinal parabolic teeth modifications are introduced in order to minimize peak to peak STE for a design load equal to F=6000 N. This gear geometry configuration is considered as the deterministic case.

In this study, tolerance ranges of profile errors, longitudinal errors and misalignments are chosen considering the quality class 7 of the AFNOR French Standard NF E23-006. This quality class is often used in industrial applications (gearbox, machine tool,…). All the geometry faults are assumed to be truncated gaussian parameters. They induce a gap between the actual teeth surfaces and theoretical teeth surfaces. This gap is expressed by two parameters $H_\alpha$ and $F_\alpha$ governing profile errors and two parameters $H_\beta$ and $F_\beta$ governing longitudinal errors (see Fig. 1). The F symbol concerns quadratic errors while H symbol concerns linear errors. These four parameters are linear combination of manufacturing errors and shaft misalignments. They are also gaussian parameters. The two first statistical moments for these parameters are given in Table 2.

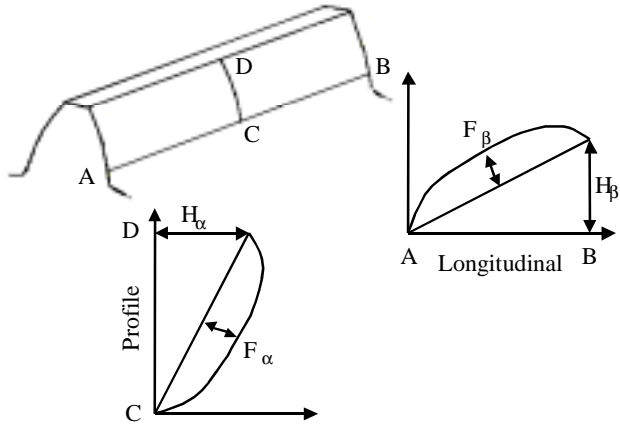

*Fig. 1 Description of used manufacturing errors.*

*Table 1 Main geometrical characteristics of spur gear.*

|  | Pinion | Driven wheel |
|---|---|---|
| **Number of teeth** | 37 | 71 |
| **Base radius (mm)** | 52.153 | 100.077 |
| **Normal module (mm)** | 3 | |
| **Pressure angle** | 20° | |
| **Facewidth (mm)** | 24 | |
| **Centre distance (mm)** | 162 | |

*Table 2 Statistical Moments of Geometry Faults.*

|  | Mean value (µm) | Standard deviation (µm) |
|---|---|---|
| $F_\alpha$ | 20 | 4.53 |
| $H_\alpha$ | 0 | 3.4 |
| $F_\beta$ | 16 | 4 |
| $H_\beta$ | 0 | 4.63 |

**Statistics of STE**

In order to compute STE time history, it is necessary to solve the static equilibrium of gear contact for each angular position of the pinion. A cinematic analysis of gear mesh allows to locate the line of contact between teeth. Equations of static equilibrium are generated from discretisation of theoretical contact lines in N nodes.

$$[H]\{P\}=\Delta\{I\}-\{e\}+\{Y\} \qquad (1)$$

$$^t\{I\}\{P\}=F \qquad (2)$$

$$\text{either } P_k=0 \text{ or } Y_k=0 \qquad (3)$$

subjected to constraints

$$P_k \geq 0,\ Y_k \geq 0 \text{ and } \Delta \geq 0. \qquad (4)$$

Unknown variables are STE expressed as a displacement $\Delta$ along the action line, the (N×1) normal load vector $\{P\}$ and the (N×1) vector of slack variables $\{Y\}$. [H] is the (N×N) non-diagonal compliance matrix computed by using FEM modelling of toothed wheel. $\{I\}$ is the (N×1) identity vector and $\{e\}$ is the (N×1) vector of initial gaps which corresponds to teeth modifications and geometrical faults. F is the input normal load. For this example, we considered rigid wheel bodies with teeth geometry described above. We have used a modified simplex algorithm described in [4] to solve these equations.

Introducing manufacturing tolerances, we used Taguchi's method to estimate the two first statistical moments of peak to peak STE $\Delta_{pp}$. The use of pertubation method to treat randomness is not adapted because equation (3) introduces non-linearity in the equation system. Results are validated by comparison with those obtained by Monte-Carlo simulations. Results are given in Table 3 for input normal loads F=6000 N and F=3000 N. The standard deviation/mean value ratio is very high for both loads. Then, we can conclude on a high dispersion of dynamic response levels. Further, STE PDF is clearly dissymmetric (Fig. 2). Finally, we observe good agreement between Taguchi's method and Monte-Carlo simulations.

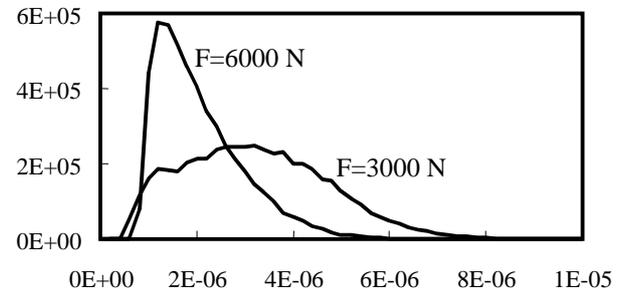

*Fig. 2 Peak to peak STE (m) PDF obtained with M.C. simulation*

*Table 3 Mean value and standard deviation obtained with rigid wheel gear bodies.*

|  | $E(\Delta_{pp})$ µm | $\sigma(\Delta_{pp})$ µm | $E(K_m)$ N/µm | $\sigma(K_m)$ N/µm |  |
|---|---|---|---|---|---|
| Monte-Carlo | 1.95 | 0.91 | 545 | 25.2 | |
| Taguchi | 1.79 | 1.06 | 547 | 26.7 | F=6000 N |
| deterministic | 0.84 |  | 557 |  | |
| Monte-Carlo | 3.11 | 1.48 | 432 | 31 | |
| Taguchi | 3.10 | 1.49 | 431 | 29.5 | F=3000 N |
| deterministic | 3 |  | 429 |  | |

**Statistics of meshing stiffness**

For each pinion angular position, the STE computation allows to estimate the meshing stiffness $k_m$ as follows :

$$k_m = \partial F/\partial \Delta \qquad (5)$$

$$k_m \approx \delta F /[\Delta(F+\delta F) - \Delta(F)] \qquad (6)$$

Time-average meshing stiffness $K_m$, which governs critical speed values, is estimated by averaging $k_m$ over a meshing period. Statistical moments of $K_m$ are obtained from Taguchi's method compared with Monte-Carlo simulations. Results for F=3000 N and F=6000 N are given in Table 3. First, mean value of $K_m$ depends strongly on the input load. Second, we observe a significant standard deviation/mean value ratio close to 7 % (respectively 5 %) for F=3000 N (respectively 6000 N) even if this ratio is smaller than the one computed for STE. Third, we notice good agreement between the two methods. At last, STE and time-average meshing stiffness are not randomly independent with a correlation coefficient close to −0.37 for the case F=3000 N. For F=6000 N, this statistical correlation is close to −0.04, so that it strongly depends on the input load F.

**Gear dynamic model and critical modes**

The dynamic model is displayed on Fig. 3. It corresponds to two rigid discs connecting by the meshing stiffness. Bearing stiffnesses $K_b$ are also introduced. The four degrees-of-freedom of this semi-definite elastic system are the two disc displacements Y parallel to the action line and the two angular positions θ.

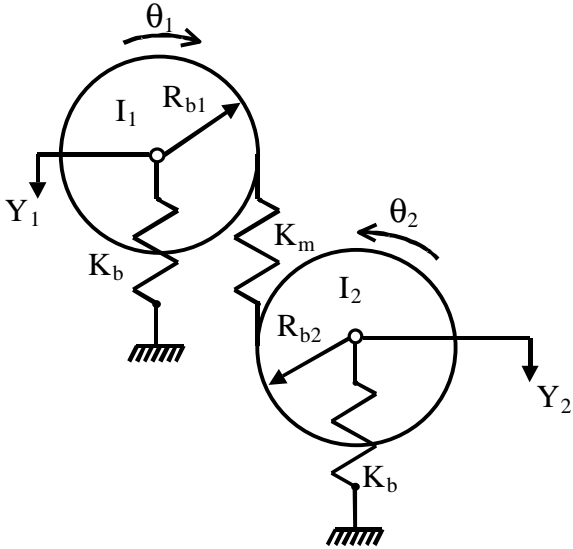

*Fig. 3 Dynamic model of the gearbox.*

The four corresponding modes are extracted analytically and the detection of the critical ones is done as follows : for each mode i, we compute the energetic coefficient defined in equation (7).

$$\rho_i = {}^t\{\phi_i\}[k_m]\{\phi_i\} / {}^t\{\phi_i\}[K]\{\phi_i\} \qquad (7)$$

where $\{\phi_i\}$ is the ith eigenvector, $[k_m]$ is the stiffness matrix including only the meshing terms and $[K]$ is the complete stiffness matrix. The higher $\rho_i$ is, the more critical the mode is [2]. For this dynamic model, only two modes (defined as modes 1 and 3) have a significant contribution ($\rho_i \neq 0$). Statistical moments of frequency $f_1$ and $f_3$ are given in Table 4. We can notice that the mean value of stochastic frequencies is different from the deterministic values without manufacturing errors. A zero order Taylor expansion would not be valid. Second, the standard deviation/mean value ratio can be higher than 2%, so that variability of geometry faults has significant effect on critical modes.

*Table 4 Mean value and standard deviation of critical eigenfrequencies.*

|  | $E(f_1)$ Hz | $\sigma(f_1)$ Hz | $E(f_3)$ Hz | $\sigma(f_3)$ Hz |  |
|---|---|---|---|---|---|
| Monte-Carlo | 1947 | 14.1 | 3314 | 48.4 | F=6000 N |
| Taguchi | 1948 | 14.9 | 3318 | 55.9 | |
| deterministic | 1954 |  | 3341 |  | |
| Monte-Carlo | 1864 | 28.2 | 3081 | 63.2 | F=3000 N |
| Taguchi | 1864 | 26.4 | 3080 | 60.4 | |
| deterministic | 1864 |  | 3075 |  | |

*Table 5 Mean value and standard deviation of energetic coefficients.*

|  | $E(\rho_1)$ | $\sigma(\rho_1)$ | $E(\rho_3)$ | $\sigma(\rho_3)$ |  |
|---|---|---|---|---|---|
| Monte-Carlo | 31.4 % | 2.4 % | 68.6 % | 2.4 % | F=6000 N |
| Taguchi | 31.2% | 2.2 % | 68.7 % | 2.2 % | |
| deterministic | 30.3 % |  | 69.7 % |  | |
| Monte-Carlo | 42.7 % | 3.7 % | 57 % | 3.7 % | F=3000 N |
| Taguchi | 42.9 % | 3.5 % | 57 % | 3.5 % | |
| deterministic | 43 % |  | 57 % |  | |

Statistical moments of $\rho_i$ are given in Table 5. Energetic repartition between modes depends on the input load. Standard deviations are equal because summation of energetic coefficients for the two modes must be 100 % for each sample. If one energetic coefficient decreases, the other one increases. Eigenfrequencies are well separated so that energetic dispersion is quite small, but an energetic crossover between the two modes is not excluded, especially for F=3000 N.

**Critical rotation speeds and associated ranges**

Critical speeds correspond to the excitation by STE of critical modes such that their eigenfrequencies correspond to meshing frequency or its harmonics :

$$n \times Z_j \times f_{rot,j} = f_i \qquad (8)$$

where n is an integer, j represents the wheel (1-input or 2-output) and i represents the critical mode (1 or 3). For a critical speed $N_j$ expressed in rpm :

$$N_j = 60 \times f_i/n \times Z_j. \qquad (9)$$

The critical speed ranges are obtained with Tchebycheff's inequality. Results are displayed in Fig. 4 for the driven wheel rotation speed (j=2) and the meshing frequency (n=1). According to Tchebycheff, critical speed has a minimum probability to be inside ranges equal to 96% for each manufactured gearbox. Range's length is about 200 rpm for the mode 1 and 500 rpm for the mode 3.

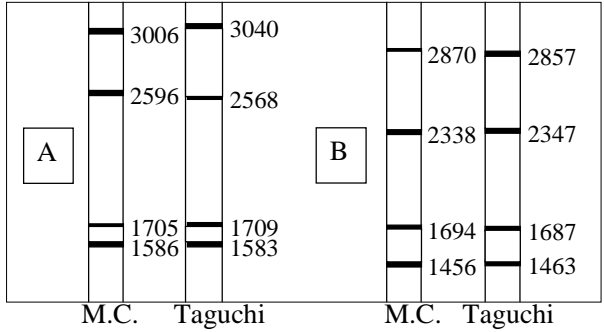

*Fig. 4 Critical output speed ranges (rpm), F=6000 N (A) and F=3000 N (B).*

## SECOND GEARBOX DYNAMIC MODEL

### Statistics of meshing stiffness

The computational method of meshing stiffness is described above. New spur gear model is introduced and consists on thin-rimmed wheels with the teeth geometry described in the first section. The computed compliance matrix [H] is considerably affected [5]. Considering the same profile and longitudinal parabolic teeth modifications, the optimal design load is close to F=3000 N. Deterministic value of time-average meshing stiffness is then $K_m$=252 N/µm.

Monte-Carlo simulations with statistical moments given in Table 2, lead to a density probability function (PDF) of $K_m$ displayed in Fig. 5. A theoretical gaussian PDF with computed first statistical moments (see Table 6) is also displayed. Meshing stiffness PDF may be approximated by a gaussian PDF, but the extension of this assumption to other input loads had to be proved. At last, Taguchi's method provides results very close to the ones obtained with Monte-Carlo simulations.

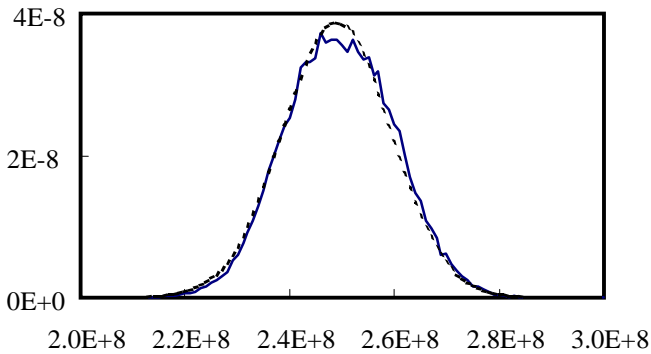

*Fig. 4 PDF of time-average meshing stiffness (N) — M.C. simulation, ---Theoretical gaussian.*

*Table 6 Mean value and standard deviation obtained with thin rimmed wheel bodies.*

|  | $E(\Delta_{pp})$ µm | $\sigma(\Delta_{pp})$ µm | $E(K_m)$ N/µm | $\sigma(K_m)$ N/µm | |
|---|---|---|---|---|---|
| Monte-Carlo | 2.23 | 1.01 | 249 | 10.3 | |
| Taguchi | 2.22 | 1.01 | 249 | 10.3 | F=3000 N |
| deterministic | 1.33 |  | 252 |  | |

### Gearbox dynamic model and critical modes

The gearbox is modelled using F.E. method (Fig. 6). The pinion and the driven wheel are modelled with concentrated masses and rotary inertia. A specific 12×12 stiffness matrix is introduced to couple the 6 d.o.f. of pinion and 6 d.o.f. of driven wheel. This matrix is defined from geometric characteristics of gear and from meshing stiffness previously computed. The shafts are discretized using beam elements with two nodes and 6 d.o.f per node. Shaft radius are equal to 45 mm and 85 mm. Distance between bearings is equal to 75 mm. The motor and the external load are connected to shafts using rotary inertia and torsion stiffness elements. A 10×10 stiffness matrix is introduced to model tapered roller bearings with use of a method described in [6]. This gearbox model contains 40 elements and 200 d.o.f..

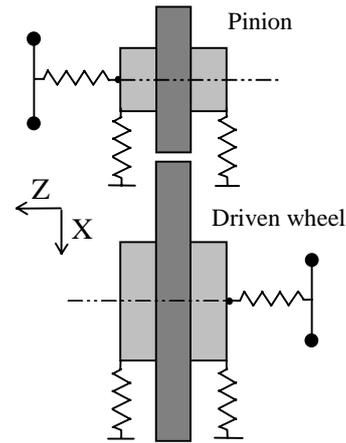

*Fig. 6 Gearbox finite element model.*

The modal basis is deduced considering the time average meshing stiffness and critical modes are extracted as above (F=3000 N). Only the 15 first modes are retained. Eigenfrequency ($f_i$) is not the only needed information. Corresponding modal shape had to be considered through energetic coefficient in order to file eigenmodes. Critical eigenfrequencies statistical moments are given in Tables 7 (Monte-Carlo) and 8 (Taguchi's method) with their corresponding energetic coefficients.

*Table 7 Mean and standard deviation of eigenmodes obtained with Monte-Carlo simulations. Critical modes are in bold letters.*

| $E(f_i)$ | $E(\rho_i)$ % | $\sigma(f_i)$ Hz | $\sigma(\rho_i)$ % |
|---|---|---|---|
| 1- 0 Hz | 0 | 0 | 0 |
| 2- 250 Hz | 3.6 | 0.2 | 0.15 |
| 3- 407 Hz | 2.6 | 0.22 | 0.1 |
| 4- 720 Hz | 0 | 0 | 0 |
| 5- 1379 Hz | 0 | 0 | 0 |
| 6- 1676 Hz | 0 | 0 | 0 |
| **7- 1758 Hz** | **32** | **12.6** | **2.88** |
| 8- 1738 Hz | 2.6 | 0.9 | 3.8 |
| 9- 2144 Hz | 0 | 0 | 0 |
| **10- 2420 Hz** | **28** | **14** | **0.7** |
| 11- 3666 Hz | 0 | 0 | 0 |
| **12- 4126 Hz** | **24** | **20** | **0.8** |
| 13- 5061 Hz | 0 | 0 | 0 |
| 14- 5169 Hz | 0 | 0 | 0 |
| **15- 5374 Hz** | **7.2** | **8** | **0.5** |

*Table 8 Mean and standard deviation of eigenmodes obtained with Taguchi's method. Critical modes are in bold letters.*

| E($f_i$) | E($\rho_i$) % | $\sigma(f_i)$ Hz | $\sigma(\rho_i)$ % |
|---|---|---|---|
| 1- 0 Hz | 0 | 0 | 0 |
| 2- 250 Hz | 3.6 | 0.17 | 0.15 |
| 3- 407 Hz | 2.6 | 0.17 | 0.1 |
| 4- 720 Hz | 0 | 0 | 0 |
| 5- 1379 Hz | 0 | 0 | 0 |
| 6- 1676 Hz | 0 | 0 | 0 |
| **7- 1758 Hz** | **31** | **12.6** | **4.3** |
| 8- 1738 Hz | 3.6 | 1 | 5.5 |
| 9- 2144 Hz | 0 | 0 | 0 |
| **10- 2420 Hz** | **28** | **14** | **0.7** |
| 11- 3666 Hz | 0 | 0 | 0 |
| **12- 4125 Hz** | **23.6** | **20** | **0.9** |
| 13- 5061 Hz | 0 | 0 | 0 |
| 14- 5169 Hz | 0 | 0 | 0 |
| **15- 5373 Hz** | **7.2** | **7.8** | **0.5** |

Three critical modes are identified (7$^{th}$, 10$^{th}$ and 12$^{th}$ modes). critical frequency variability is not high. The Taguchi's method gives accurate results close the ones provided by Monte-Carlo simulations excepted for two neighbours modes (7$^{th}$ and 8$^{th}$ modes). Contrary to the deterministic case, mean value of 7$^{th}$ critical natural frequency is higher than mean value of 8$^{th}$ natural frequency. An energetic cross-over and a frequency cross-over between modes are observed. Standard deviation of ($\rho_i$) is higher than its mean value for 7$^{th}$ mode. Taguchi's method does not provide accurate results in this particular case.

**Critical speed ranges**

For input speed varying from 0 to 5000 rpm, critical speed ranges are obtained from Tchebycheff's inequality and displayed on Fig. 7. According to Tchebycheff, critical input speed has a minimum probability to be inside ranges equal to 96% for each manufactured gearbox. They correspond to excitation in a resonant manner of 7$^{th}$ and 10$^{th}$ modes by the meshing frequency (n=1) and at its first harmonic (n=2). Range's length is about 200 rpm for the 7$^{th}$ and the 10$^{th}$ modes.

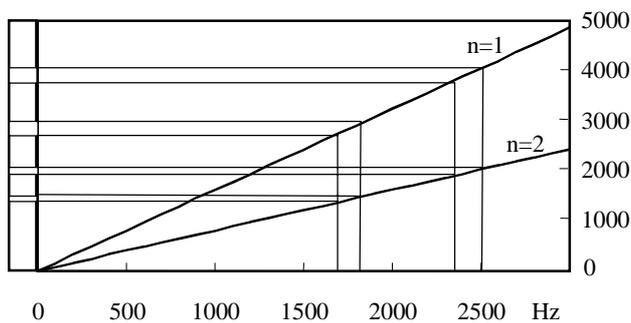

*Fig. 7 Two main critical input speed ranges (r.p.m.) for excitation at meshing frequency (n=1) and at first harmonic (n=2).*

As a confirmation of these predicted critical ranges, we show one dynamic response obtained for a possible teeth geometry configuration (Fig. 8). This dynamic response is obtained after solving the gearbox differential motion equation with a spectral and iterative method [7].

It corresponds to dynamic meshing force versus input speed. Main resonant phenomena are observed for input speeds close to 3000 rpm and 4000 rpm. Other secondary phenomena are observed for input speeds close to 2000 rpm and 1500 rpm. They are due to excitation of critical modes by the second harmonic of meshing frequency. All these rotation speeds are included in the predicted critical ranges.

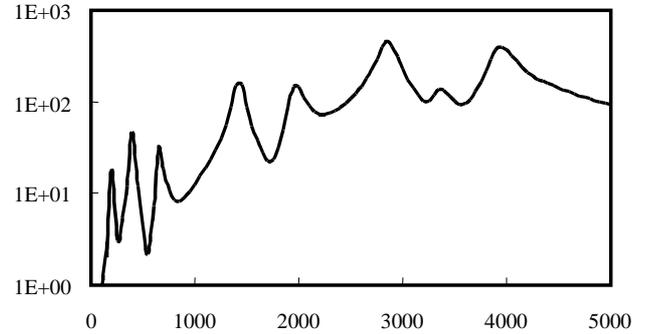

*Fig. 8 Dynamic meshing force (N) versus input speed (rpm).*

**CONCLUSION**

For gearbox manufactured in large number, we have quantified excitation levels dispersion and the stochastic frequency ranges where resonant phenomena may occur. The dispersion values show that it is necessary to control gear-manufacturing tolerances to reduce mean value of radiated noise as well as its variability.

The statistical Taguchi's method offers a powerful and simple tool for treating gear dynamical behaviour with manufacturing tolerances. In most of treated cases, the prediction of statistical moments is close to the ones given by Monte-Carlo simulationss. The computation time required by Taguchi's method is 300 times smaller than the one required for Monte-Carlo simulation with 30000 samples. This methodology had to be applied on more complex systems having a lot of energetic modes.